# Trivial topological phase of CaAgP and the topological nodal-line transition in CaAg(P$_{1-x}$As$_x$)

N. Xu[1*], Y. T. Qian[2,3,4,5], Q. S. Wu[2,3], G. Autès[2,3], C. E. Matt[6], B. Q. Lv[4,5,6], M. Y. Yao[6], V. N. Strocov[6], E. Pomjakushina[7], K. Conder[7], N. C. Plumb[6], M. Radovic[6], O. V. Yazyev[2,3], T. Qian[4], H. Ding[4,5,8], J. Mesot[2,6,9] and M. Shi[6,*]

[1] *The Institute of Advanced Studies, Wuhan University, Wuhan 430072, China*

[2] *Institute of Physics, École Polytechnique Fédérale de Lausanne, CH-1015 Lausanne, Switzerland*

[3] *National Centre for Computational Design and Discovery of Novel Materials MARVEL, Ecole Polytechnique Fédérale de Lausanne (EPFL), CH-1015 Lausanne, Switzerland*

[4] *Beijing National Laboratory for Condensed Matter Physics and Institute of Physics, Chinese Academy of Sciences, Beijing 100190, China*

[5] *University of Chinese Academy of Sciences, Beijing 100049, China*

[6] *Swiss Light Source, Paul Scherrer Institut, CH-5232 Villigen PSI, Switzerland*

[7] *Laboratory for Multiscale Materials Experiments, Paul Scherrer Institut, CH-5232 Villigen, Switzerland*

[8] *Collaborative Innovation Center of Quantum Matter, Beijing, China*

[9] *Laboratory for Solid State Physics, ETH Zürich, CH-8093 Zürich, Switzerland*

\* E-mail: nxu@whu.edu.cn, ming.shi@psi.ch

**Abstract**

By performing angle-resolved photoemission spectroscopy and first-principle calculations, we address the topological phase of CaAgP and investigate the topological phase transition in CaAg(P$_{1-x}$As$_x$). We unveil that in CaAgP the bulk band gap and surface states with a large band-width are topologically trivial, in agreement with hybrid DFT calculations. The calculations also indicate that application of 'negative' hydrostatic pressure can transform trivial semiconducting CaAgP into an ideal topological nodal-line semimetal phase. The topological transition can be realized by partial isovalent P/As substitution at $x = 0.38$.

**Introduction**

In the last decade, the band topology theory has achieved a series of spectacular successes, leading to the theoretical prediction and experimental realization of two-dimensional (2D) and three-dimensional (3D) topological insulators [1-2], as well as to the understanding of the topological phase transition [3], which goes beyond the theory of phase transitions associated with symmetry breaking. Recent advances in the field have extended the topological classification to gapless systems [4-6], i.e. topological semimetals, in which new types of fermionic quasiparticles are excited near the touching points of the bulk bands in the vicinity of the Fermi level. Similar to topological insulators, the surface states associated with the nontrivial bulk electronic structure in a topological semimetal are robust against disorder and impurities. Furthermore, it is expected that the low energy excitations in the bulk states could result in pronounced anomalies in magnetotransport properties [7-9].

Through identifying the finger-prints of their bulk and surface electronic structures, Dirac, Weyl and multiple-point semimetals, in which bands touch at discrete points in the momentum space, have been realized in Na$_3$Bi and Cd$_3$As$_2$ [10-17], the TaAs [18-35], WTe$_2$ [36-40] and WP$_2$ family [41-42], and the MoP family [43-50] materials. However, a less common situation is when the band gap closes along one-dimensional lines in the momentum space, forming nodal-line [51-58], nodal-chain [59-60] or nodal-net [61-63] topological band degeneracies. In the systems that gaps close along one-dimensional lines novel quantum phenomena are expected to emerge, such as the flat "drumhead" surface bands, long-range Coulomb interactions [64], a quasi-topological electromagnetic response [65], heavy-electron behavior [66] and even the high-$T_c$ superconductivity [67]. The experimental results on PbTaSe$_2$ and ZrSiS have provided a solid evidence of the Dirac-like features [68-70] and topological surface states [71]. However, the electronic structure of these candidate materials is rather complex with additional trivial Fermi surfaces (FS) contributing to the transport properties.

Distinct from all other candidates, hexagonal pnictides CaAgP and CaAgAs (see the crystal structure in Fig. 1a-b) are suggested as ideal topological nodal-line semimetals (TNLSs), whose FS contains only the Dirac nodal-line in the bulk Brillouin zone (BZ) (Fig. 1d) [72]. Therefore, the transport properties in CaAgP and CaAgAs would originate predominantly from the electronic states near the bulk nodal-line with drumhead surface states contribution, which makes the system as an ideal platform for studying the intrinsic topological physics of TNLSs. Here, we present a comprehensive investigation on the surface and bulk electronic structures of CaAgP by combining angle-resolved photoemission spectroscopy (ARPES) and first-principles calculations. A hole-like 2D surface band is observed in vacuum ultra-violate (VUV-) ARPES measurements. Different from the expected drumhead surface band in nodal-line semimetals, the observed surface band with a large Fermi velocity is attributed to the surface resonance band, which is consistent with our calculations. The soft X-ray (SX-) ARPES measurements on the 3D bulk electronic states further indicate that the electronic structure of CaAgP is topological trivial with an energy gap between valence and conduction bands. We verified that the first-principle calculations based on density functional theory using the Heyd-Scuseria-Ernzerhof (HSE) hybrid exchange-correlation functional can render a correct description of the topological feature for the bands near the $E_F$, and well reproduce the electronic structure determined from the ARPES measurements on CaAgP. On the other hand, our experimental results are qualitatively different to that from the calculations using Perdew–Burke-Ernzerhof (PBE) exchange functional, which predict CaAgP to be a topological semimetal or insulator [72]. We further simulated the pressure effect in CaAgP by isotropically scaling the lattice parameters, and found that a topological nodal-line transition can be induced by applying 'negative' hydrostatic pressure. The unusual reverse pressure effect on the semiconductor-topological semimetal transition in CaAgP is due to a competition mechanism between the decrease of both the band width and the Ag($s$)-P($p$) hybridization gap upon increasing lattice parameters. In realistic situations, such 'negative' hydrastatic pressure can be realized by isovalent substitution of P with As in CaAg(P$_{1-x}$As$_x$), which results in isotropic expansion of the lattice, and leads to a topological phase transition at $x$ = 0.38. Our results indicate that CaAg(P$_{1-x}$As$_x$) is a promising candidate system for realizing the novel nodal-line semimetal states with the simplest nodal-line FS, as well as for studying the corresponding topological phase transition.

**Method**

**Sample synthesis.**

Single crystals of CaAgP were grown by a chemical vapour transport in a temperature gradient 1000 => 950C, using 0.5 g of polycrystalline CaAgP as a source and iodine as a transport agent with a concentration of 5 mg/cm$^3$. Polycrystalline CaAgP was synthesized by a solid-state reaction using elemental calcium, silver and red phosphorus of a minimum purity 99.99%. The respective amounts of silver and red phosphorus were mixed and pressed into a pellet in the He-glove box and placed into alumina crucible together with a corresponding amount of metallic calcium. The crucible was annealed in the evacuated quartz ampule at 1100C for 20 h. The laboratory X-ray diffraction measurements, which were done at room temperature using Cu K$\alpha$ radiation on Brucker D8 diffractometer, have proven that the obtained crystals are single phase with the hexagonal structure of space group $P\bar{6}2m$.

**Angle-resolved photoemission spectroscopy**

Clean surfaces for ARPES measurements were obtained by cleaving CaAgP samples *in situ* in a vacuum better than 5 × 10$^{-11}$ Torr. VUV-ARPES measurements were performed at the Surface and Interface (SIS) beamline at the Swiss Light Source (SLS) with a Scienta R4000 analyser and 'Dreamline' beamline of the Shanghai Synchrotron Radiation Facility (SSRF) with a Scienta Omicron DA30L analyser. Soft x-ray ARPES measurements were performed at the Advanced Resonant Spectroscopies (ADRESS) beamline at SLS with a SPECS analyser [73], and data were collected at $T$ = 10 K using circular-polarized light with an overall energy resolution on the order of 50-80 meV for FS mapping and 40-60 meV for high-resolution cuts measurements.

**Calculation Methods**

Our PBE calculations on CaAgP and CaAgAs performed using the WIEN2k code within generalized gradient approximation (GGA) [74] showed that CaAgP is a nodal-line semimetal, in agreement with the previous study [72]. The calculations performed using the Vienna ab-initio simulation package (VASP) used both the GGA as well as Heyd–Scuseria–Ernzerhof (HSE) [75-76] screened hybrid functional. A planewave cutoff energy of 300 eV and a 6 x 6 x 9 $k$-point mesh were used during the self-consistent field procedure. The Wannier90 package [77] was utilized for generating the Wannier-function based tight-binding (WFTB) model Hamiltonian with initial projectors of Ag $5s$, Ag $4d$, P(As) $3s$ and P(As) $3p$ atomic orbitals. WannierTools packaged [78] was used for calculating the surface-state spectra. The lattice constants of $a = b = 7.1131$ Å, $c = 4.1944$ Å for CaAgP and $a = b = 7.2947$ Å, $c = 4.3007$ Å for CaAgAs were adapted from experiments [79]. The energy gap of $CaAg(As_xP_{1-x})$ was obtained from the mixed WFTB Hamiltonian constructed by assuming that its matrix elements are the linear interpolation of that for stoichiometric compounds CaAgP and CaAgAs. Such linear interpolation of Hamiltonian matrix elements can be considered as a virtual crystal approximation [80].

**Results**

**The predication of TNLS states in CaAgP and CaAgAs**

CaAgP has a ZrNiAl-type hexagonal structure belonging to space group P$6\bar{2}$m, the lattice parameters are $a = b = 7.04$ Å and $c = 4.17$ Å. As shown in Fig. 1a, in this system the mirror symmetry about the Ca plane between the two layers of AgP$_4$ tetrahedra is preserved, but inversion symmetry is broken. AgP$_4$ tetrahedra form a 3D kagome-triangular lattice in which Ca atoms are located at the shared edges and corners (Fig. 1b). The CaAgP single crystals synthesized by chemical transport method have irregular shapes with millimeter sizes (Fig. 1c). The bulk and ($\bar{1}$100) surface BZs are shown in Figure 1c. In the previous study [72] carried out using the PBE exchange-correlation functional, it was suggested that CaAgP hosts an ideal TNLS state, in which FS contains only the nodal line (red circle in Fig. 1d) formed by the touching points of the valence and conduction bands (Fig. 1e-g). In CaAgAs, the isovalent substitution of P by As leads to an isotropic expansion, resulting in increased lattice parameters $a = b = 7.20$ Å and $c = 4.27$ Å, which preserve the same $c/a$ ratio as in CaAsP [79]. At the same time, in CaAgAs the enhanced spin-orbit coupling (SOC) opens a narrow gap at the band crossing points, which drive the system from TNLS to a topological insulator with narrow band gap.

**Trivial surface states on the ($\bar{1}100$) surface of CaAgP**

One manifestation of the TNLS is a flat drumhead surface band that occurs in the interior of the surface projection of the nodal line and contributes substantially to the density of states near the Fermi level. Using surface sensitive VUV-ARPES the expected surface states should be observed directly, and distinguished from the bulk states. Figure 2a shows the FS map, acquired with photon energies in the range of $h\nu$ = 30-120 eV, in the $k_x$-$k_y$ plane of the momentum space. It should be noted that the ARPES data were acquired from the ($\bar{1}100$) surface and $k_y$ is the out-of-plane component in the ARPES measurements. Around the $\bar{\Gamma}$ point there is a sharp band (α band) (shown with red arrows in Fig. 2a) whose Fermi momenta do not vary with the change of the photon energy, indicating its 2D character and surface origin. The α band is hole-like and forms a nephroid-shape Fermi surface around the surface BZ center (Fig. 2b) with Fermi velocity about 2.7 eV/Å and 5.4 eV/Å along the $\bar{\Gamma} - \bar{X}$ and $\bar{\Gamma} - \bar{A}$ lines, respectively (Fig. 2c-d). Distinct from the topological drumhead surface band expected in a TNLS, which should only appear inside the surface projection of nodal-lines, the surface band observed here occurs in a wide region of surface BZ and has a large band-width. From the consistency with surface state calculation shown in the right part of Fig. 2b-d, we attribute this surface band as a topologically trivial surface resonance band. On the other hand, only with the VUV-ARPES results themselves one cannot exclude the possible TNLS state in CaAgP because a large shift of the surface chemical potential due to impurities or other imperfections could result in the flat drumhead states becoming unoccupied and inaccessible in the ARPES experiments. The trivial topological property of CaAgP is unambiguously demonstrated by the bulk electronic state with a semiconducting gap determined by SX-ARPES and consistent DFT+HSE calculations that is described as following.

**Trivial bulk state of CaAgP with a band gap between valence and conduction**

**bands**

To further investigate the topological nature of CaAgP we applied SX-ARPES to determine the bulk band structure, which is more bulk sensitive and has better out-of-plane momentum resolution [81]. As shown in Fig. 3a, the FS acquired in the photon energy range of 350 - 650 eV repeats periodically in the bulk BZs in both in-plane ($k_x$) and out-of-plane ($k_y$) directions, thus represents the FS of bulk states. Figure 3b shows the FS map in the Γ-M-L-A plane, acquired with photon energy $h\nu$ = 580 eV. In Fig. 3c we plot the band dispersions along the high symmetry line Γ-K-M (cut1 in Fig. 3a) in a large energy scale. The measured energy bands are directly compared with that calculated using PBE (black lines) and HSE (blue lines) functionals with chemical potential shifts downward by 0 and 0.08 eV, respectively. The tiny adjustment of the chemical potential in the comparison indicates that the experimentally determined small FS volume is very close to that from calculations and suggests additional charge carrier concentration induced by impurities or other imperfections is very low. The ARPES spectra near $E_F$ along Γ-M and Γ-A (cut2 and cut3 in Fig. 3b) are shown in Fig. 3d-e, respectively. The corresponding curvature plots and energy distribution curves are displayed in Fig. 3f-g and Fig. 3h-i, respectively, to better visualize the band dispersions. The bulk band structure determined in the SX-ARPES experiment is qualitatively different to that from PBE calculations, which predicts that CaAgP is a TNLS in the previous studies. In early PBE calculations the valence and conduction bands overlap, forming a toroidal Fermi surface when the chemical potential is slightly away from the touching point, and both electron- and hole-like bands appear in the Γ-M direction. However, in our ARPES experiment, no electron-like band is observed, as shown in Fig. 3d and Fig. 3f. Another fingerprint of the TNLS state is that the hole-like band maximum locates at different energies in the Γ-M and Γ-A directions, with the energy difference equal to the overlap width between valence and conduction bands (Δ in Fig. 3d and 3e). The top of the observed hole-like band along the Γ-A direction is located at a much higher energy compared to the PBE prediction (Fig. 3e). In Fig. 3j we show the band dispersions along Γ-M (green circles) and Γ-A (blue crosses) near $E_F$, obtained by fitting the peak positions of the momentum distribution curves, and corresponding fitting with simple parabola (green and blue lines). The energy positions of band top along Γ-M and Γ-A show a negligible difference, instead of Δ = 400 meV as suggested by PBE calculations. On the other hand, the observed bulk band structure is in good agreement with our HSE hybrid functional calculations, as shown in Fig. 3f and 3g. The current ARPES and HSE calculations provide an accurate description of the electronic structure of CaAgP, which is topologically trivial with an energy gap between the valence and conduction bands.

**'Negative' hydrostatic pressure and isovalent As substitution induces topological nodal-line transition in CaAgP**

Using the HSE methodology, which provides accurate description of the band structure of CaAgP, we carried out a thorough investigation on topological properties induced by the hydrostatic pressure modeled by isotropically scaling the lattice parameters with a factor of $\varepsilon = a/a_0 = b/b_0 = c/c_0$. Remarkably, the band gap between the conduction and valence bands, $E_c - E_v$, decreases with increasing the lattice parameters ($\varepsilon > 1$), that correspond to negative hydrostatic pressure, and closes at the critical point $\varepsilon_c = 1.024$. At this point a Dirac semimetal state is realized. The band gap changes the sign at $\varepsilon > \varepsilon_c$ driving CaAgP into the TNLS phase (Fig. 4a). In contrast to the usual pressure effect when positive pressure ($\varepsilon < 1$) results in a larger band-width and could induce an insulator-metal transition, the insulator-topological semimetal transition in CaAgP takes place at the negative hydrostatic pressure. This effect in CaAgP is related to the reduction of the hybridization gap, that is the gap induced by the hybridization between the Ag $s$ atomic orbitals and P(As) $p$ atomic orbitals, decreases upon the increase of lattice parameters. The gap eventually vanishes and the system undergoes a semiconductor to TNLS transition. To best of our knowledge, this is the first example showing the 'reversal' of hydrostatic pressure driving semiconductor-(topological) semimetal transition.

The "negative" hydrostatic pressure effect can be realized upon partial or full substitution of P with isoelectronic As in CaAg(P$_{1-x}$As$_x$) as the $c/a$ ratio is practically preserved in such replacement. To study the evolution of the electronic structure with the As substitution in detail, we constructed a Wannier function Hamiltonians for the interpolation between CaAgP and CaAgAs using the HSE functional. We found that upon the substitution of P with As the conduction and valence bands approach each other, touch at the CaAg(P$_{0.62}$As$_{0.38}$) composition, and realize an inverted band structure in CaAgAs (Fig. 4b). The energy gap vanishes at the critical point $x_c = 0.38$ marking a transition from the topological trivial phase to the TNLS phase takes place (Fig. 4a). We notice, however, that more delocalized As 4*p* atomic orbitals (compared to P 3*p* atomic orbitals) further amend in closing the band gap, compare to the pure negative hydrostatic pressure effect. We also notice that when SOC is included in the calculation, a tiny gap (70 meV) opens at the nodal line of CaAgAs, driving the system into a narrow gap topological insulator, similar to the results of previous studies with PBE. Unfortunately, a number of hole-like carriers induced by disorder or impurity states exist in the CaAgAs single crystal samples, which shift the chemical potential dramatically [82-86]. Further improvements of sample quality are required.

## Summary

Using a combination of ARPES measurements and first-principle calculations we demonstrate that CaAgP exhibits a topologically trivial phase. The experimentally determined electronic structure can be well described by DFT calculations using the HSE functional incorporating spin-orbit coupling. We further investigated the effect of pressure on the electronic structure of CaAgP finding that the system transform into an ideal TNLS phase by applying a negative hydrostatic pressure that corresponds to a lattice parameters scaling factor above $\varepsilon_c = 1.024$. In practice the negative hydrostatic pressure can be realized by means of isovalent substitution of P with As in CaAg($P_x As_{1-x}$), with the predicted topological critical point $x_c = 0.38$ at which a Dirac semimetal phase occurs. The SOC at high levels of As substitution open a small energy gap at the nodal line driving the system into the strong topological insulator regime.

**Acknowledgements**

This work was supported by NCCR MARVEL funded by the Swiss National

Science Foundation, the Swiss National Science Foundation (Grant No. 200021-159678), the Sino-Swiss Science and Technology Cooperation (Grant No. IZLCZ2-170075), "the Fundamental Research Funds for the Central Universities" (Grant No. 2042018kf-0030). N.X. acknowledges support by the Thousand Talents Plan and Wuhan University start up funding. Y.Q. acknowledges support by a MARVEL INSPIRE Potentials Master's Fellowship. First-principles calculations were performed at the Swiss National Supercomputing Centre (CSCS) under project s675 and the facilities of Scientific IT and Application Support Center of EPFL.

**Competing financial interests**

The authors declare no competing financial interests.

# Figures

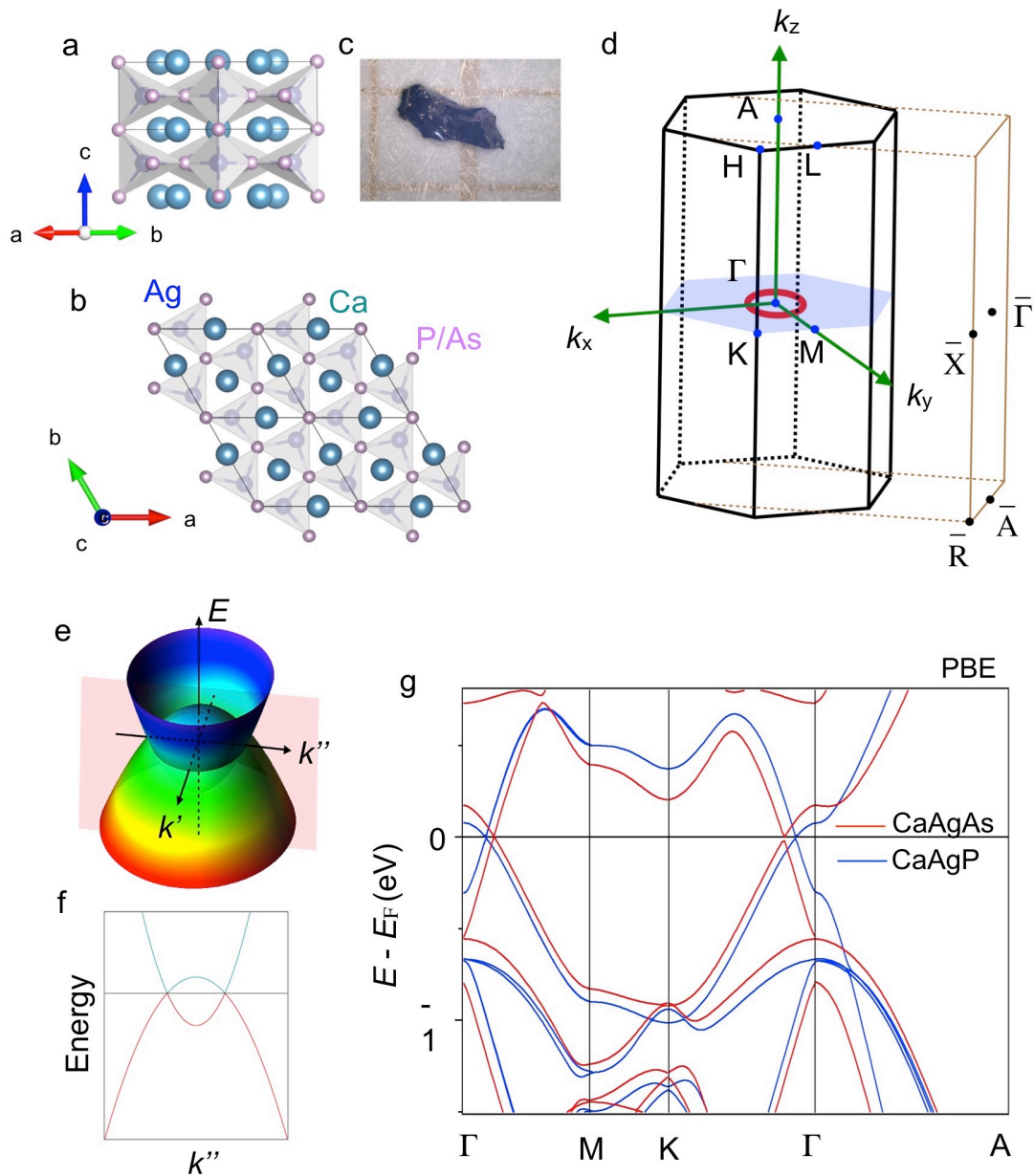

**Figure 1. Topological nodel-line semimetal phase in CaAgP/CaAgAs. a-b**, Side and top views of the crystal structure of CaAgP/CaAgAs. **c,** The picture of a typical piece of CaAgP single crystal used in ARPES experiments, taken under the microscope on a millimeter scale. **d**, Bulk and surface BZs with high-symmetry points labelled. **e-f**, Illustration of the band dispersions near the nodal line in the $k_x$-$k_y$ plane. **g**, The bulk electronic structure of CaAgP and CaAgAs from PBE calculations.

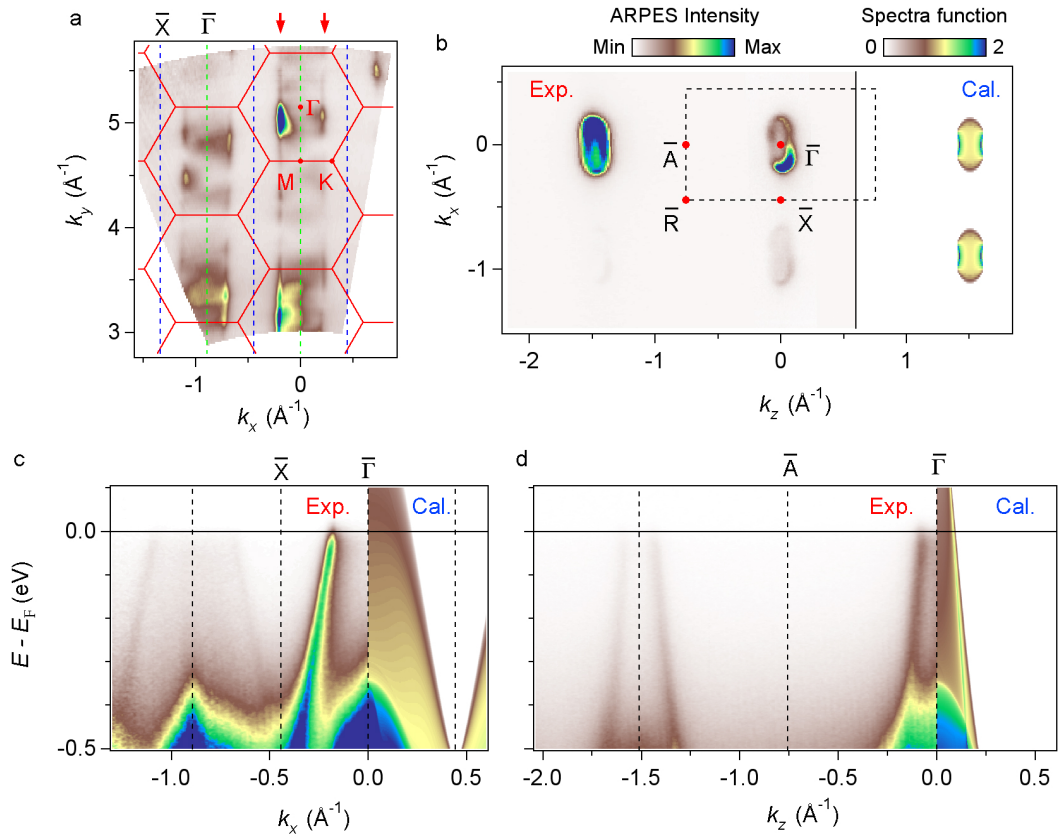

**Figure 2. Surface states at the ($\bar{1}$100) surface of CaAgP. a**, The FS acquired with VUV ARPES in the photon energy range from 30 eV to 120 eV. $k_x$ ($k_y$) is the in-plane (out-of-plane) component of the crystal momentum. Red arrows indicate the FS of 2D surface states. **b**, The FS in the surface BZ, obtained from the ARPES experiment (left) and calculation (right). **c-d**, Band structure along the $\bar{\Gamma} - \bar{X}$ and $\bar{\Gamma} - \bar{A}$ directions, obtained from the ARPES experiment (left) and HSE calculations (right).

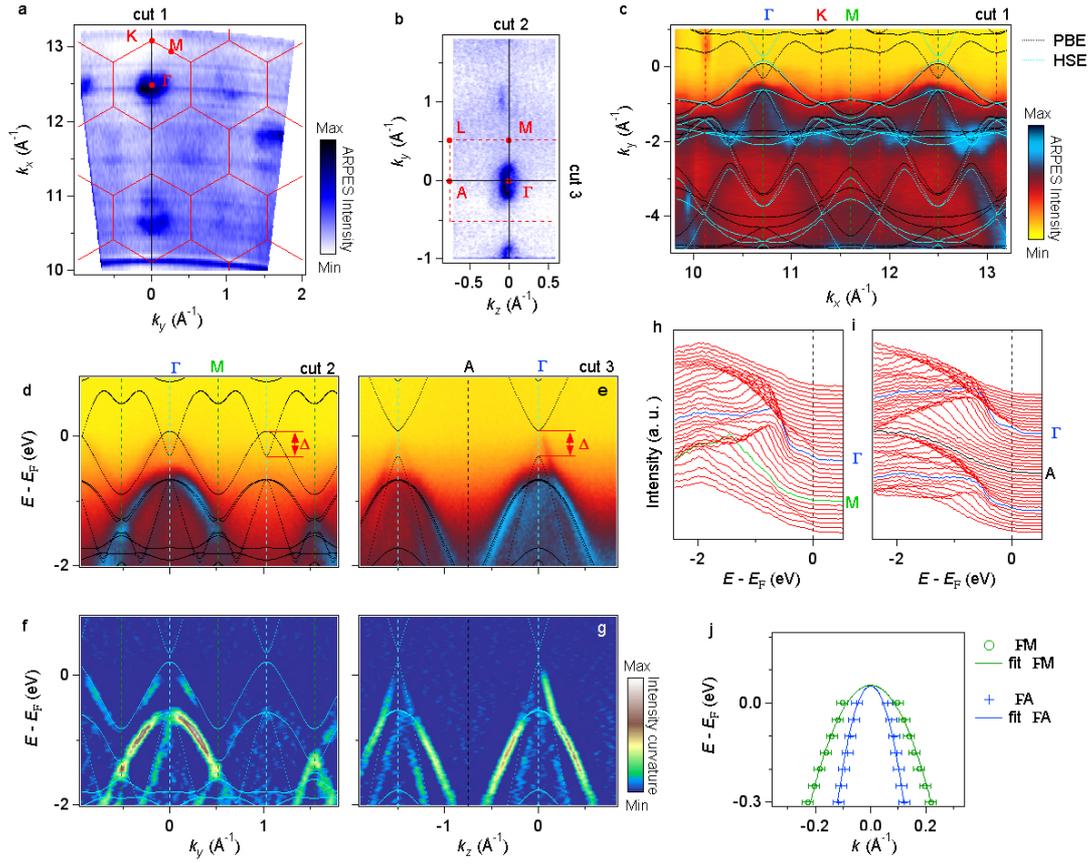

**Figure 3. The bulk band structure of CaAgP. a**, The FS in the $k_x$-$k_y$ plane of the bulk BZ, acquired with SX-ARPES in the photon energy range from 350 eV to 650 eV. **b**, The FS recorded with $h\nu$ = 580 eV. **c**, ARPES spectra along the high symmetry line Γ-K-M (cut1 in **a**), the band structures obtained from PBE and HSE calculations are overlaid on the spectra. **d**, ARPES spectra along Γ-M and the electronic structure from PBE calculations. **e**, The curvature plots of ARPES spectra in **d**, overlaid with the band structure from HSE calculations. **f-g**, Same as **d-e**, but along Γ-A direction. **h-i**, Energy distribution curves along the Γ-M and Γ-A directions, respectively. **j**, The extracted and fitting hole-like band dispersion along Γ-M and Γ-A directions.

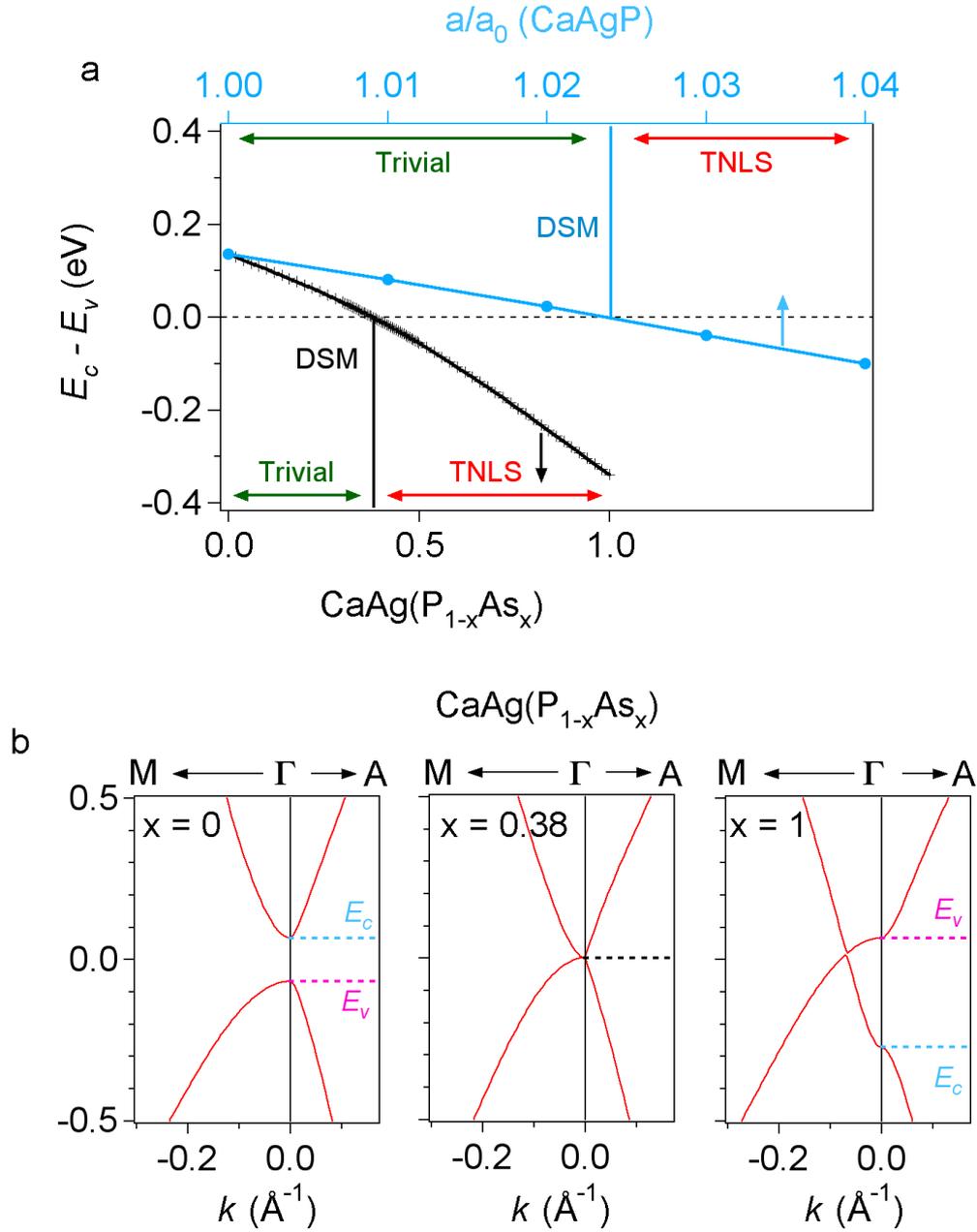

**Figure 4. Simulations of the hydrostatic pressure effect and isovalent As substitution in CaAgP. a**, The topological phase diagram with hydraulic pressure and isovalent As substitution in CaAgP. The energy difference between conduction and valence bands ($E_c - E_v$) induced by isovalent As substitution is extracted from the Wannier functions interpolations of CaAgP and CaAgAs within the HSE calculations. The hydrostatic pressure effect in CaAgP is simulated by isotropically scaling lattice parameter $\varepsilon = a/a_0 = b/b_0 = c/c_0$. **b**, Calculated band structures near $E_F$ by using HSE functional for $CaAg(P_{1-x}As_x)$ with $x = 0, 0.38$ and 1, respectively.